# Fuzzy Clustering of Web Documents Using Equivalence Relations and Fuzzy Hierarchical Clustering

Satendra Kumar[1], Mamta Kathuria[2], Alok Kumar Gupta[3], Monika Rani[4]

*Abstract*-**The conventional clustering algorithms have difficulties in handling the challenges posed by the collection of natural data which is often vague and uncertain. Fuzzy clustering methods have the potential to manage such situations efficiently. Fuzzy clustering method is offered to construct clusters with uncertain boundaries and allows that one object belongs to one or more clusters with some membership degree. In this paper, an algorithm and experimental results are presented for fuzzy clustering of web documents using equivalence relations and fuzzy hierarchical clustering.**

## I. INTRODUCTION

Clustering is an unsupervised learning problem; it deals with finding a structure in a collection of unlabeled data. A loose definition of clustering could be the process of organizing objects into groups whose members are similar in some way. A cluster is therefore a collection of objects which are similar between them and are dissimilar to the objects belonging to other clusters. Clustering algorithms can have different properties:

- Hierarchical or flat: Hierarchical algorithms induce a hierarchy of clusters of decreasing generality, for flat algorithms, all clusters are the same.
- Iterative: The algorithm starts with initial set of clusters and improves them by reassigning instances to clusters.
- Hard and soft: Hard clustering assigns each instance to exactly one cluster. Soft clustering assigns each instance a probability of belonging to a cluster.
- Disjunctive: Instances can be part of more than one cluster.

Clustering involves two forms of object cluster which are overlap and non-overlap cluster. The cluster is said as overlap when the object can belong to other cluster and non-overlap is the object belongs to one and only cluster. In fuzzy clustering, data elements can belong to more than one cluster, and associated with each element is a set of membership levels. These indicate the strength of the association between that data element and a particular cluster. Fuzzy clustering is a process of assigning these membership levels, and then using them to assign data elements to one or more clusters. There are few types of fuzzy clustering, such as Fuzzy c-varieties (FCV) algorithm, adaptive fuzzy clustering (AFC) algorithm, Fuzzy C-mean (FCM) algorithm, Gustafson-Kessel (GK) algorithm and Gath-Geva (GG) algorithm. Borgelt (2003) has used FCM in the high number of dimensions and the special distribution characteristics of the data. The selection of Fuzzy methods is to find fuzzy clusters of ellipsoidal shape and differing size since the data is too complex. As the studies from Borgelt, it is said that FCM succeeds in cluster the document and yields the best result in classification accuracy.

In IR field, cluster analysis has been used to create groups of documents with the goal of improving the efficiency and effectiveness of retrieval, or to determine the structure of the literature of a field. The terms in a document collection can also be clustered to show their relationships. The two main types of cluster analysis methods are the non-hierarchical, which divide a data set of N items into M clusters and hierarchical clustering produces a nested data set in which pairs of items or clusters are successively linked. The hierarchical methods have usually been favored for cluster-based document retrieval. The commonly used hierarchical methods, such as single link, complete link, group average link, and Ward's method have high space and time requirements.

Web Mining has fuzzy characteristics, so fuzzy clustering is sometimes better suitable for Web Mining in comparison with conventional clustering. Fuzzy clustering is a technique for information retrieval. As a document might be relevant to multiple queries, this document should be given in the corresponding response sets, otherwise, the users would not be aware of it. Fuzzy clustering is a natural technique for document categorization. There are two basic methods of fuzzy clustering, one which is based on fuzzy c-partitions, is called a fuzzy c-means clustering method and the other, based on the fuzzy equivalence relations, is called a fuzzy equivalence clustering method. The purpose of this research is to propose a search methodology that consists of how to find relevant information from WWW. In this paper, a method is being proposed of document clustering, which is based on fuzzy equivalence relation that helps information retrieval in the terms of time and relevant information [8].

[2]Mamta Kathuria is Assistant Professor in Department of Computer Science & Engineering,YMCA University of Science and Technology, Faridabad-121006, India
(e-mail: mamtakathuria7@rediffmail.com)
[1,3,4]Satendra Kumar, Alok Kumar Gupta and Monika Rani are M.Tech. Students in Department of Computer Science & Engineering, YMCA University of Science and Technology, Faridabad-121006, India
(e-mail:satendra04cs41@gmail.com, alokgupta.niec@gmail.com, monikarani1988@gmail.com)

Rest of the paper is organized as follows. Section 2 presented the related work. The fuzzy clustering algorithm has been described in Section 3. Section 4 shows the experimental result, how to do fuzzy clustering of documents. Section 5 presents the conclusion.

## II. RELATED WORK

The goal of traditional clustering is to assign each data point to only one cluster. In contrast, Fuzzy clustering assigns different degrees of membership to each point where the membership of a point is shared among various clusters (Fung, 2001). Article that characterize a given knowledge domain are somehow associated with each other. Those articles may also be related to articles of other domains. Hence, documents may contain information that is relevant to different domains to some degree. With Fuzzy clustering methods documents are attributed to several clusters simultaneously and thus, useful relationships between domains may be uncovered, which would otherwise be neglected by hard clustering methods.

Data Mining has emerged as a new discipline in world of increasingly massive datasets. Data Mining is the process of extracting or mining knowledge from data. Data Mining is becoming an increasingly important tool to transform data into information. Knowledge Discovery from Data i.e. KDD is synonym for Data Mining [8].

Friedman Menahem et al. in [1] have been presented a methodology for a new Fuzzy-based Document Clustering Method (FDCM), to cluster documents that are represented by variable length vectors. Each vector element consists of two fields. The first is an identification of a key phrase (its name) in the document and the second denotes a frequency associated with this key phrase within the particular document.

Jursic Matjaz et al. in [2] have been presented the fuzzy clustering of 2-dimensional points and documents. For the needs of documents clustering they implemented fuzzy c-means in the Text Garden environment.

Oren Etzioni was the person who coined the term Web Mining first time. Initially two different approaches were taken for defining Web Mining. First was a "process-centric view", which defined Web Mining as a sequence of different processes [6] whereas second was a "data-centric view", which defined Web Mining in terms of the type of data that was being used in the mining process [7]. The second definition has become more acceptable, as is evident from the approach adopted in most research papers [3][5]. Web Mining is also a cross point of database, information retrieval and artificial intelligence [4].

## III. PROPOSED WORK

We are proposed a fuzzy clustering method which is based upon fuzzy equivalence relation. Clustering documents with keywords stored in system at any location. We extract all the words from the entire set of documents and eliminate stop words such as 'a' 'the' 'in' 'on' 'and' etc. from all the documents. There is a proposed algorithm for fuzzy clustering of web documents using equivalence relation.

**Proposed Algorithm:**

1. Input the documents files.
2. Remove the stop words from all document files.
3. List of keywords (common words) with document id is generated.
4. Each keyword is assigned a keyword id.
5. Mapping the keyword id and document id to generate document clustering data on the basis of same keyword.
6. Determine a fuzzy compatibility relation in terms of an appropriate distance function applied on the given data.

$$R(x_i, x_k) = 1 - \delta \left( \sum_{J=1}^{p} | x_{ij} - x_{kj} |^q \right)^{1/q} \ldots \ldots (i)$$

For all pairs $(x_i, x_k) \in X$(set of data), $q \in RT$ and $\delta \rightarrow$ constant that ensures that $R(x_i, x_k) \in [0,1]$, Clearly, $\delta$ is the inverse value of the largest distance in X.

7. Calculate transitive closure of this fuzzy compatibility relation. Calculated a relation R(X,X), and its transitive closure RT (X,X) can be determined by simple algorithm that consists of the following three steps:

    i. R' = R U (R o R)

    ii. If R' ≠ R, make R = R' and go to step 1

    iii. Stop R' = $R_T$

8. Calculate α-cut of transitive closure relation.
9. Draw the dendogram with help of α-cut. It will show the clustering of web documents. For the value of q=1 or 2 it should be same because it will be apply on same data.

## IV. EXPERIMENTAL RESULTS

We present the results of an experimental evaluation of the fuzzy clustering technique using equivalence and fuzzy clustering. We implemented our proposed algorithm in Java and run our experiments on NetBeans IDE 7.1.2 over 64-bit Windows7 Operating System, Intel Core i5@ 2.40GHz with 4GB of RAM. In our system the experimental result shows the same clusters of web documents after applying q=1 or 2.Here we show the experimental results for values q=2 and q=1, experimental results generate the same clusters of web documents.

(a) Input the folder of documents files and enter the value of q=2 then start the process and generate the keywords with document Ids.

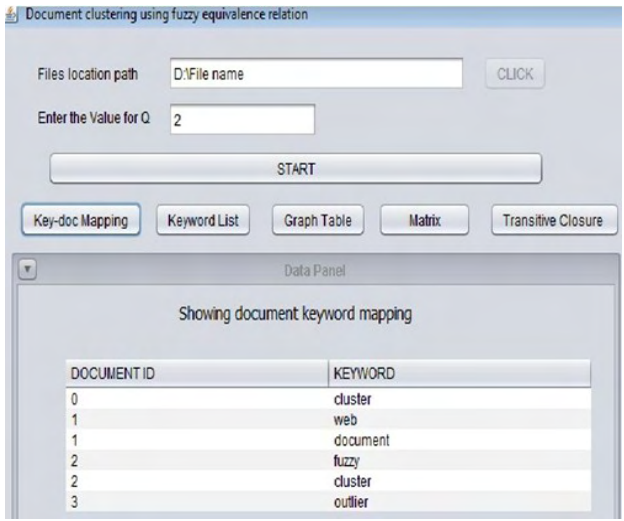

Fig.1.Input of files and keyword with doc id

In the above fig. there are four documents and five keywords. By applying above algorithm, analyze the data for q= 2.which is corresponds to the Euclidean distance.

(b) Keyword Id is generated for each keyword and every keyword should be occurred at once.

Fig.2. Keyword with keyword ID

(c) Showing mapping data table of above tables (a) and (b).

Showing KeywordID and DocumentID Mapping Data Table

| x1 | x2 | x3 | x4 | x5 | x6 |
|---|---|---|---|---|---|
| 1 | 2 | 3 | 4 | 5 | 6 |
| 0 | 1 | 1 | 2 | 2 | 3 |
| 0 | 1 | 2 | 3 | 0 | 4 |

Fig.3. Document clustering data

There are six data points:- **x1= (0,0) , x2= (1,1) ,x3=(1,2) , x4=(2,3), x5= (2,0),x6=(3,4).** The largest Euclidean distance between any pair of given data points is 5 then δ = 1/5 =0.20.

(d) Showing matrix of Fuzzy Compatibility relation. Now calculate membership values of R for equation (i)
$R(x_2, x_4) = 1 - 0.20(1^2 + 2^2)^{0.5} = 0.55$
When determined, relation R may conveniently be represented by the matrix for data points.

Showing Matrix after Fuzzy Compatibility relation

| x1 | x2 | x3 | x4 | x5 | x6 |
|---|---|---|---|---|---|
| 1.0 | 0.72 | 0.55 | 0.28 | 0.6 | 0.0 |
| 0.72 | 1.0 | 0.8 | 0.55 | 0.72 | 0.28 |
| 0.55 | 0.8 | 1.0 | 0.72 | 0.55 | 0.43 |
| 0.28 | 0.55 | 0.72 | 1.0 | 0.4 | 0.72 |
| 0.6 | 0.72 | 0.55 | 0.4 | 1.0 | 0.18 |
| 0.0 | 0.28 | 0.43 | 0.72 | 0.18 | 1.0 |

Fig.4. Fuzzy relation matrix

(e) Calculating the transitive closure of above relation matrix. This relation is not max-min transitive. It is transitive closure. This relation includes three distinct partitions of its α - cuts:

Showing Transitive Closure Matrix

| x1 | x2 | x3 | x4 | x5 | x6 |
|---|---|---|---|---|---|
| 1.0 | 0.72 | 0.72 | 0.72 | 0.72 | 0.72 |
| 0.72 | 1.0 | 0.8 | 0.72 | 0.72 | 0.72 |
| 0.72 | 0.8 | 1.0 | 0.72 | 0.72 | 0.72 |
| 0.72 | 0.72 | 0.72 | 1.0 | 0.72 | 0.72 |
| 0.72 | 0.72 | 0.72 | 0.72 | 1.0 | 0.72 |
| 0.72 | 0.72 | 0.72 | 0.72 | 0.72 | 1.0 |

Fig.5.shows transitive closure matrix

(f) Calculate the α-cut of above transitive closure matrix.

## Showing Alpha Cuts

| Alpha cuts | Members |
|---|---|
| [0.0, 0.72] | {{X1,X2,X3,X4,X5,X6,}} |
| (0.72, 0.8] | {{X2,X3,},{X1},{X4},{X5},{X6}} |
| (0.8, 1.0] | {{X1,}{X2,}{X3,}{X4,}{X5,}{X6,}} |

Fig.6. α-cuts of transitive closure matrix

**(g)** Draw the dendogram with help of α-cut. It will show the clustering of web documents. For the value of q=1 or 2 it should be same because it will be apply on same data.
(For q=2)

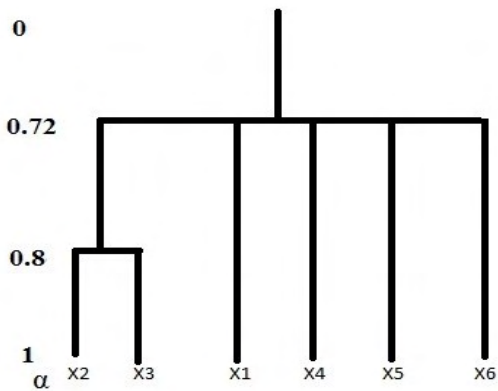

Fig.7. Dendogram representation

Now we check the experimental result for q = 1 in eq. (i) which represents the Hamming distance. Since the largest hamming distance in the data is 7, we have δ=0.14.

**(a)** The matrix form of relation R is given by eq. (i) is now

## Showing Matrix after Fuzzy Compatibility relation

| x1 | x2 | x3 | x4 | x5 | x6 |
|---|---|---|---|---|---|
| 1.0 | 0.71 | 0.57 | 0.29 | 0.71 | 0.0 |
| 0.71 | 1.0 | 0.86 | 0.57 | 0.71 | 0.29 |
| 0.57 | 0.86 | 1.0 | 0.71 | 0.57 | 0.43 |
| 0.29 | 0.57 | 0.71 | 1.0 | 0.57 | 0.71 |
| 0.71 | 0.71 | 0.57 | 0.57 | 1.0 | 0.29 |
| 0.0 | 0.29 | 0.43 | 0.71 | 0.29 | 1.0 |

Fig.8. Fuzzy compatibility matrix for q=1

**(b)** Its transitive closure is shown below.

## Showing Transitive Closure Matrix

| | x1 | x2 | x3 | x4 | x5 | x6 |
|---|---|---|---|---|---|---|
| | 1.0 | 0.71 | 0.71 | 0.71 | 0.71 | 0.71 |
| | 0.71 | 1.0 | 0.86 | 0.71 | 0.71 | 0.71 |
| | 0.71 | 0.86 | 1.0 | 0.71 | 0.71 | 0.71 |
| | 0.71 | 0.71 | 0.71 | 1.0 | 0.71 | 0.71 |
| | 0.71 | 0.71 | 0.71 | 0.71 | 1.0 | 0.71 |
| | 0.71 | 0.71 | 0.71 | 0.71 | 0.71 | 1.0 |

Fig.9. Transitive closure matrix

**(c)** The transitive closure relation gives the following portions in its α – cuts

## Showing Alpha Cuts

| Alpha cuts | Members |
|---|---|
| [0.0, 0.71] | {{X1,X2,X3,X4,X5,X6,}} |
| (0.71, 0.86] | {{X2,X3,},{X1},{X4},{X5},{X6}} |
| (0.86, 1.0] | {{X1,}{X2,}{X3,}{X4,}{X5,}{X6,}} |

Fig.10. α-cuts of transitive closure matrix

**(d)** Draw the dendogram with help of α-cut. It will show the clustering of web documents.

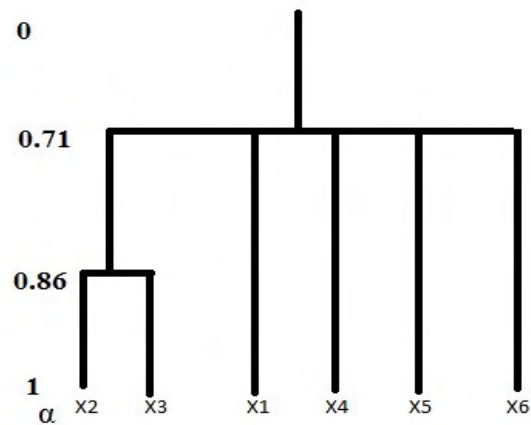

Fig.11. Dendogram representation

Here dendogram for both values (1, 2) of q is similar such we say that our experimental result is correct. The dendrogram is a graphical representation of the results of hierarchical cluster analysis. This is a tree-like plot where each step of hierarchical clustering is represented as a fusion of two branches of the tree into a single one. The branches represent clusters obtained on each step of hierarchical clustering.

## V. CONCLUSION AND FUTURE WORK

Fuzzy clustering is better than conventional clustering because it is suitable for Web Mining. Fuzzy clustering is also useful to detect the outlier data point or documents. This proposed technique for document clustering, based on fuzzy logic approach improves relevancy factor because experimental results shows the same fuzzy hierarchical clustering for both values of q. In the proposed algorithm there is a fuzzy compatibility equation (i) is useful to calculate the membership values which shows the belonging status of the web documents to each other. If the value of q is one it calculates membership values for hamming distance. If the value of q is two then it calculates the membership values for the Euclidean distance. This technique keeps the related documents in the same cluster so that searching of documents becomes more efficient in terms of time complexity. In future work we can also improve the relevancy factor of fuzzy clustering to retrieve the web documents.